\documentclass[%
 aip,
 amsmath,amssymb,
 reprint,%
]{revtex4-1}

\usepackage{graphicx}% Include figure files
\usepackage{dcolumn}% Align table columns on decimal point
\usepackage{bm}% bold math
\usepackage[utf8]{inputenc}
\usepackage[T1]{fontenc}
\usepackage{mathptmx}
\usepackage{color}
\usepackage{appendix}

\usepackage{caption}
\usepackage{subcaption}

\begin{document}

\preprint{AIP/123-QED}

\title[]{
An N\textsuperscript{5}-scaling excited-state-specific perturbation theory
}
% Force line breaks with \\

\author{Rachel Clune}
\affiliation{
Department of Chemistry, University of California, Berkeley, California 94720, USA 
}

\author{Jacqueline A. R. Shea}
\affiliation{
Department of Chemistry, University of California, Berkeley, California 94720, USA 
}

\author{Eric Neuscamman}
\email{eneuscamman@berkeley.edu}
\affiliation{
Department of Chemistry, University of California, Berkeley, California 94720, USA 
}
\affiliation{Chemical Sciences Division, Lawrence Berkeley National Laboratory, Berkeley, CA, 94720, USA}

\date{\today}% It is always \today, today,

\begin{abstract}
We show that by working in a basis similar to that of the natural transition
orbitals and using a modified zeroth order Hamiltonian, the cost of a
recently-introduced perturbative correction to excited state mean field theory
can be reduced from seventh to fifth order in the system size.
The (occupied)$^2$(virtual)$^3$ asymptotic scaling matches that
of ground state second order M{\o}ller-Plesset theory,
but with a significantly higher prefactor because the bottleneck
is iterative: it appears in the  Krylov-subspace-based solution of the
linear equation that yields the first order wave function.
Here we discuss the details of the modified zeroth order Hamiltonian
we use to reduce the cost as well as the automatic code generation process
we used to derive and verify the cost scaling of the different terms.
Overall, we find that our modifications have little impact on the method's
accuracy, which remains competitive with singles and doubles
equation-of-motion coupled cluster.
\end{abstract}

\maketitle

\section{Introduction}
\label{sec:intro}

%\textcolor{red}{
%--- Something about the Par. ---\\
%}

Although mean field methods like Hartree-Fock (HF) theory often
succeed in making qualitatively correct predictions about how electrons
distribute themselves within a molecule, making quantitative
energetic predictions at the precision necessary to aid in designing
and interpreting experiments usually requires grappling with the
finer-grained wave function details that arise from electron correlation.
In many contexts, especially when considering ground states in closed-shell
molecules, density functional theory (DFT) fills this role at a relatively
low computational expense.
However, even in these DFT-friendly systems, there are areas
-- such as the treatment of weak intermolecular interactions
\cite{grimme2016dispersion} --
where more expensive wave-function-based methods remain essential
to, for example, help choose which empirical functional to trust.
In electronically excited states, open-shell character is the norm, and
in practice DFT faces serious challenges and is less predictive than in
ground states.
These challenges include both the inability of time-dependent DFT (TD-DFT) to
relax the shapes of orbitals not directly involved in the excitation
\cite{Ziegler2009,Ziegler2015,Zhao2019}
and the tendency of self-consistent DFT, as used for example in the restricted
open-shell Kohn Sham (ROKS) \cite{Shaik1999,Kowalczyk2013} method,
to over-delocalize \cite{Yang2008Localization,isborn2013CT}
unpaired electrons or holes.
Although the latter issue can be mitigated by using hybrid
and range-separated functionals, \cite{Tozer2013tuning}
it nonetheless persists. \cite{Zhao2020}
If wave-function-based methods are to help make up for DFT's difficulties in this
area, it is highly desirable that they overcome these challenges while retaining
electron correlation corrections that are as computationally affordable as possible.
In this study, we take a step in this direction by reformulating a second-order
perturbative correction to excited state mean field (ESMF) theory
\cite{Shea2018,Shea2020GVP,Zhao2020}
so that its asymptotic cost scaling can reach parity with
its ground state counterpart.

In the world of closed-shell ground states, the simplest and usually the most
affordable approach to electron correlation aside from DFT is second-order
M{\o}ller-Plesset perturbation theory (MP2). \cite{MollerPlesset1934}
In a canonical implementation, the asymptotic scaling of this method is
$N_o^2N_v^3$, where $N_o$ and $N_v$ denote the number of occupied and virtual orbitals
in the HF reference, respectively. \cite{headgordon1999semidirectMP2}
Note that, for simplicity, we will throughout this paper consider $N_v$ to be
interchangeable with $N$, the total number of molecular orbitals, when
discussing asymptotic scaling.
Although significantly higher than the cost scaling of many widely used
density functionals, the cost of MP2 is significantly lower than
the sixth-order cost of coupled cluster theory with singles and doubles (CCSD),
positioning it as the least expensive wave-function-based ground state correlation
method in wide use.
The excited-state-specific ESMP2 theory \cite{Shea2018} that we focus on
in this study was designed to closely mirror MP2 theory, correcting ESMF
in the same way that MP2 corrects HF, achieving rigorous size intensivity,
and working in an uncontracted first order interacting space.
Unfortunately, the fact that the ESMF reference already contains single
excitations means that this interacting space now includes both the doubles
and triples excitations.
Acting the zeroth order Hamiltonian in this space thus involves contracting
a two-index Fock operator with a six-index amplitude tensor, leading to
seventh order scaling with the system size and a theory that is decidedly
less practical than the ground state theory that it seeks to mimic.

To overcome this difficulty, we exploit the fact that a wave function
that is a linear combination of singles excitations, such as configuration
interaction singles (CIS), can be written as a sum of just $N_o$
configuration state functions (CSFs) under a particular occupied-occupied
and virtual-virtual rotation of the orbital basis.
Working in this basis --- which for CIS itself is the natural transition
orbital basis \cite{Martin_2003_NTO} but for ESMF will be slightly different
due to its excited-state-specific orbital relaxations ---
the coulomb operator no longer connects the singly excited reference
function to the whole triples space.
Separating the triples in to those that connect with the reference and those that do not,
one expects the unconnected triples (which are by far the larger group)
to be less important, and so a more aggressive approximation of the zeroth
order Hamiltonian in that space is somewhat justified.
In particular, we will approximate the Fock operator in the unconnected
space by its diagonal
(note that, unlike for a HF reference, the Fock operator
derived from the ESMF one-body density matrix is not diagonal)
at which point the unconnected triples no longer contribute to the theory at
all, as they have no direct connection to the reference through the coulomb
operator and no connection to the first order wave function through
the zeroth order Hamiltonian.
As we will discuss, this step immediately drops the scaling to sixth order.
To drop the scaling further, we note that, in the vast majority
of low-lying excitations in weakly correlated molecules, only a
small number out of the $N_o$ singly excited CSFs in the ESMF reference
are expected to have large coefficients.
By extending the diagonal approximation of the zeroth order Hamiltonian
into the space of all triples that only connect to small parts of the
reference, a reduction to $N_o^2N_v^3$ cost scaling is achieved.
Again, as these triples are less important, this approximation is not
expected to make much difference, and indeed this expectation is confirmed
by a comparison to results from the seventh-order parent method.
Thus, by working in a particular orbital basis and slightly modifying
the zeroth order Hamiltonian, the cost scaling of ESMP2 can be
brought in line with that of MP2, even if the prefactor remains higher
due to the off-diagonal zeroth order Hamiltonian and thus a need to
iteratively solve a linear equation.

Although ESMP2, in either its original or its more efficient form,
is similar to a number of other excited state perturbation theories,
it also possesses important differences.
When compared to CIS(D), which uses the HF orbitals and derives
its triples amplitudes from the ground state MP2 doubles, \cite{head1994cis_d}
ESMP2 is instead wholly excited-state-specific:
the orbitals are relaxed variationally for the excited
state at the ESMF level, and the triples are derived via
the excited state's first order wave function equation.
In comparison to the recently-introduced
driven similarity renormalization group
VCIS-DSRG-PT2 approach, \cite{li2017VCISDRPT2}
ESMP2 again enjoys orbitals that are relaxed for the excited state,
and it does not require the choice of an active space,
making it easier to apply in a black-box manner.
% Finally, in contrast with CASPT2,
Finally, in contrast with complete active space second order
perturbation theory (CASPT2),
\cite{roos1982caspt2,andersson1992caspt2}
N-electron valence perturbation theory (NEVPT2),
\cite{angeli2001NEVPT2} and VCIS-DSRG-PT2,
ESMP2 sticks to an uncontracted and thus orthonormal first order
interacting space, which circumvents the need to address
the potential for linear dependencies. 
That said, ESMP2 has much in common with CASPT2,
and as we will see in the results, often hews rather closely
to CASPT2 when it comes to predicting excitation energies.
Again, ESMP2 achieves this without using an
active space, which offers significant simplicity at the cost
of being inappropriate for strongly correlated systems.

This paper is organized as follows.
We begin by discussing the ESMF reference and how it
can be simplified by working in a particular orbital basis,
after which we discuss the first order wave function
and the newly-modified zeroth order Hamiltonian.
We then briefly discuss the automated approach we
employ for term derivation and code generation,
which allows us to make a detailed investigation of
each term's scaling, the outcomes of which we present
in the first subsection of the results.
We then delve into the method's accuracy, first in
a set of small molecules that are mostly single-CSF
in character and then in a collection of ring
excitations, in which multi-CSF character is more prevalent.
We end our results section with an explicit test of
size intensivity before 
concluding with a summary and a brief discussion of
possible future directions.

\section{Theory}
\label{sec:theory}

\subsection{Zeroth Order Wave Function}
\label{sec:zeroth_order_wfn}

To simplify our implementation, we have chosen to work with a slightly
simplified version of the ESMF ansatz
\begin{align}
\left|\Psi_0\right> = e^{\hat{X}}
%\Bigg( 
% c_0|\Phi\rangle + 
\sum_{ia} 
C_{ia} \big(
\hat{a}_{a\uparrow}^+ \hat{a}_{i\uparrow}
% \left|\Phi\right> 
\pm 
% \tau_{ia}
\hat{a}_{a\downarrow}^+ \hat{a}_{i\downarrow} \big)
\left|\Phi\right>
%\Bigg)
\label{eqn:psi0}
\end{align}
in which we have set the coefficient on the un-excited closed-shell
reference determinant $\left|\Phi\right>$ to zero.
This simplification avoids a significant number of terms in the
perturbation theory, but it does mean that we are assuming
that the closed shell determinant is unimportant in the excited
state, which is not universally true.
%The absence means that the excited states studied in this paper are purposefully chosen such that this component would be negligible for these states. 
Here the $\pm$ sign is plus (minus) for singlet (triplet) states,
$\bm{C}$ is the matrix of single-excitation configuration interaction (CI)
coefficients, $\hat{X}$ is an anti-Hermitian one-electron operator responsible
for excited-state-specific orbital relaxations,
and we adopt the convention of referring to occupied and unoccupied (virtual)
orbitals in $\left|\Phi\right>$ by the indices $i$,$j$,$k$,$l$
and $a$,$b$,$c$,$d$, respectively.
After relaxing $\hat{X}$ and $\bm{C}$ to find the energy stationary point
corresponding to the excited state in question (which may for example
proceed by guessing the CIS wave function and applying a generalized
variational principle \cite{Shea2020GVP}),
we take a singular value decomposition of the rectangular matrix $\bm{C}$
\begin{align}
    \bm{C} = \bm{U} \bm{\Lambda} \bm{V}^{+}
\end{align}
where, if we assume that there are more virtual than occupied orbitals,
$\bm{\Lambda}$ is the $N_o \times N_o$ diagonal matrix of singular values.
Now, note that the Hamiltonian can be transformed into an
orbital basis that eliminates $\bm{U}$ and $\bm{V}$ and thus renders
the reference wave function in a particularly simple form.
\begin{align}
   \hat{H}^{\mathrm{HF}} \hspace{0mm} \rightarrow & \hspace{2mm}
   e^{-\hat{Z}} e^{-\hat{Y}} e^{-\hat{X}}
   \hat{H}^{\mathrm{HF}}
   e^{ \hat{X}} e^{ \hat{Y}} e^{ \hat{Z}}
\end{align}
Here we have started in the HF orbital basis (as indicated by the Hamiltonian
$\hat{H}^{\mathrm{HF}}$), rotated via $\hat{X}$ into the ESMF orbital basis,
and then rotated via the one-electron anti-Hermitian operators 
$\hat{Y}$ and $\hat{Z}$,
which perform occupied-occupied and virtual-virtual rotations, respectively.
The $\hat{Y}$ rotation can be chosen so as to eliminate $\bm{U}$,
and likewise the $\hat{Z}$ rotation can be used to eliminate
$\bm{V}$, leaving us with a greatly simplified CI expansion
\begin{align}
    \left|\Psi_0\right> \hspace{1mm} \rightarrow & \hspace{2mm}
    \sum_{i} \Lambda_{ii}
    \big(
          \hat{\sigma}_{i\uparrow}^+ \hat{\tau}_{i\uparrow}
          \pm 
          \hat{\sigma}_{i\downarrow}^+ \hat{\tau}_{i\downarrow}
    \big)
    \left|\Phi\right>
    \label{eqn:psi0_tp}
\end{align}
involving a sum over the singular values of $\bm{C}$.
The corresponding virtual-orbital creation operators $\hat{\sigma}^+$
and occupied-orbital destruction operators $\hat{\tau}$ now come in pairs,
one for each occupied orbital.
We refer to each of these pairs as a transition orbital pair (TOP), and
note that, if the optimal ESMF orbitals were the same as the RHF orbitals,
the TOPs would be equivalent to the natural transition orbital (NTO) pairs.
\cite{Martin_2003_NTO}
It is important to emphasize that Eqs.\ (\ref{eqn:psi0}) and (\ref{eqn:psi0_tp})
refer to exactly the same zeroth order wave function, they simply express it
in different orbital bases.
We now turn to the definition of our first order wave function,
where the TOP basis will allow for useful groupings of the
triples excitations into separate categories that we will exploit
in order to achieve a lower asymptotic cost scaling.

\subsection{First Order Wave Function}
\label{sec:first_order_wfn}

To begin, let us specify the language we will use for describing
excitations as well as the the different orbital labels that
we employ when working in the TOP orbital basis.
First, throughout this paper, we will refer to excitation levels
relative to the closed shell.
In this language, our reference is a superposition
of single excitations, and our first-order interacting space
consists of double and triple excitations.
As for how we label orbitals, let us adopt an orbital ordering
in which the spatial orbitals are numbered 1 through $N$.
In addition to the occupied orbitals with destruction operators
$\hat{\tau}_i$ (with $i$ allowed to range from 1 through $N_o$) and the
corresponding TOP virtual orbitals whose creation operators are
$\hat{\sigma}^+_a$ (with $a$ allowed to range from $N_o+1$ through $2N_o$),
there are additional virtual orbitals (AVOs),
whose creation operators we will denote by $\hat{\nu}^+_a$
(with $a$ allowed to range from $2N_o+1$ through $N$).
When necessary, we will denote virtual orbitals that may be either TOP virtuals
or AVOs using the creation operators $\hat{w}^+_a$, where $a$ can range
from $N_o+1$ through $N$.
Finally, when we denote a TOP virtual orbital using an occupied index,
as for example in the operator $\hat{\sigma}^+_{i\uparrow}$ in
Eq.\ (\ref{eqn:psi0_tp}), this implies the TOP virtual orbital
with index $a=i+N_o$ that is the partner of the $i$th occupied orbital
in the TOP orbital basis representation of $\left|\Psi_0\right>$.

With these orbital definitions in hand and working in the TOP orbital basis,
we now point out that while the Hamiltonian, through its two-electron part,
can connect the singly-excited wave function $\left|\Psi_0\right>$
to the full space of doubly excited determinants, it only connects
$\left|\Psi_0\right>$ to a subset of the triply excited determinants.
In particular, the matrix element
\begin{align}
    H_{ijk}^{abc} =
    \left<\Psi_0\right|
    \hat{\tau}^+_k
    \hat{\tau}^+_j
    \hat{\tau}^+_i
    \hat{w}_a
    \hat{w}_b
    \hat{w}_c
    \hat{H}
    \left|\Psi_0\right>
    \label{eqn:triples_element}
\end{align}
will only be nonzero if there is at least one TOP amongst the occupied and
virtual orbitals $i$,$j$,$k$,$a$,$b$,$c$.
Put another way, this matrix element is zero if $d \ne l+N_o$ for all
$d\in\{a,b,c\}$ and $l\in\{i,j,k\}$, as a nonzero element
is only possible if one of the three excitations was
already present in $\left|\Psi_0\right>$, and $\left|\Psi_0\right>$
only contains TOP excitations.
In contrast, this matrix element can be nonzero if $d = l+N_o$ for at
least one $d$,$l$ pair from $d\in\{a,b,c\}$ and $l\in\{i,j,k\}$.
Thus, in our first order wave function
\begin{align}
    \left|\Psi_1\right> = & \hphantom{+}
    \sum_{ijab} T_{ij}^{ab}
    \hat{w}^+_{a\uparrow}
    \hat{w}^+_{b\uparrow}
    \hat{\tau}_{j\uparrow}
    \hat{\tau}_{i\uparrow}
    \left|\Phi\right>
    \notag \\
    & +
    \sum_{ijab} T_{ij}^{ab}
    \hat{w}^+_{a\downarrow}
    \hat{w}^+_{b\downarrow}
    \hat{\tau}_{j\downarrow}
    \hat{\tau}_{i\downarrow}
    \left|\Phi\right>
    \notag \\
    & +
    \sum_{ijab} S_{ij}^{ab}
    \hat{w}^+_{a\uparrow}
    \hat{w}^+_{b\downarrow}
    \hat{\tau}_{j\downarrow}
    \hat{\tau}_{i\uparrow}
    \left|\Phi\right>
    \notag \\
    & +
    \sum_{ijkabc} T_{ijk}^{abc}
    \hat{w}^+_{a\uparrow}
    \hat{w}^+_{b\uparrow}
    \hat{w}^+_{c\uparrow}
    \hat{\tau}_{k\uparrow}
    \hat{\tau}_{j\uparrow}
    \hat{\tau}_{i\uparrow}
    \left|\Phi\right>
    \notag \\
    & +
    \sum_{ijkabc} T_{ijk}^{abc}
    \hat{w}^+_{a\downarrow}
    \hat{w}^+_{b\downarrow}
    \hat{w}^+_{c\downarrow}
    \hat{\tau}_{k\downarrow}
    \hat{\tau}_{j\downarrow}
    \hat{\tau}_{i\downarrow}
    \left|\Phi\right>
    \notag \\
    & +
    \sum_{ijkabc} S_{ijk}^{abc}
    \hat{w}^+_{a\uparrow}
    \hat{w}^+_{b\downarrow}
    \hat{w}^+_{c\downarrow}
    \hat{\tau}_{k\downarrow}
    \hat{\tau}_{j\downarrow}
    \hat{\tau}_{i\uparrow}
    \left|\Phi\right>
    \notag \\
    & +
    \sum_{ijkabc} S_{ijk}^{abc}
    \hat{w}^+_{a\downarrow}
    \hat{w}^+_{b\uparrow}
    \hat{w}^+_{c\uparrow}
    \hat{\tau}_{k\uparrow}
    \hat{\tau}_{j\uparrow}
    \hat{\tau}_{i\downarrow}
    \left|\Phi\right>
    \label{eqn:psi1}
\end{align}
we set to zero the values of all same-spin ($T_{ijk}^{abc}$)
and mixed-spin ($S_{ijk}^{abc}$) 
triples coefficients whose indices do not contain at least one TOP.
As we will choose our zeroth order Hamiltonian to be diagonal
in the space of triples excitations that contain no TOPs
(which we define as the N-triples space),
setting these coefficients to zero is not an approximation, but
merely the natural consequence of their Eq.\ (\ref{eqn:triples_element})
matrix elements being zero and $\hat{H}_0$ not connecting
them to any other parts of $\left|\Psi_1\right>$.
Instead, the new approximation, and the key difference from our previous
$N^7$-scaling excited-state-specific perturbation theory, \cite{Shea2018}
comes in the definition of $\hat{H}_0$, to which we now turn
our attention.

\subsection{Zeroth Order Hamiltonian}
\label{sec:zeroth_order_ham}

\begin{figure}[t]
\centering
\includegraphics[width=8.0cm,angle=0,scale=1.0]{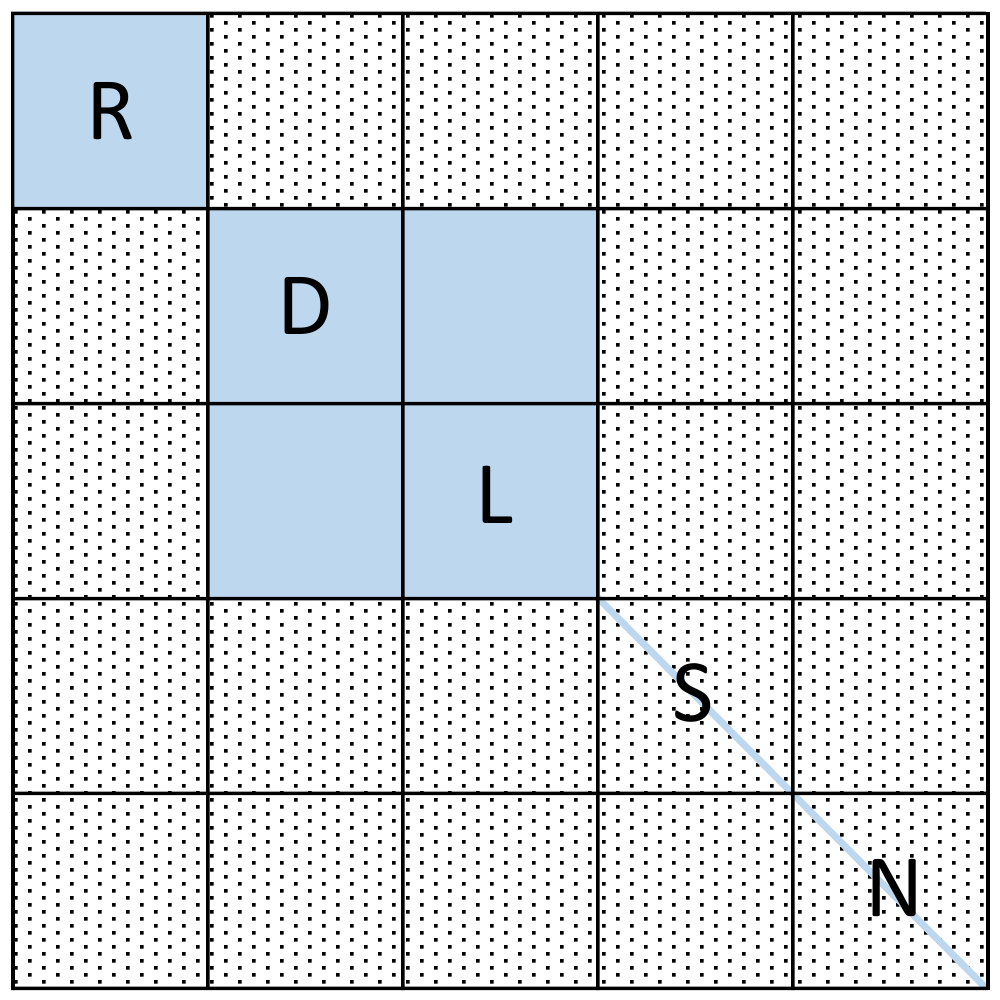}
\caption{Block structure of the zeroth order Hamiltonian.
         The matrix is zero in the dotted regions and non-zero in
         the blue regions, including on the diagonal of blocks
         S and N.
         The blocks are labeled as R for $\left|\Psi_0\right>$,
         D for double excitations,
         L for triple excitations containing at least one TOP with
         a large singular value,
         S for all other triple excitations containing at least one TOP,
         and N for the triple excitations that contain no TOPs.
         Note that no singly excited states are included in our
         first-order interacting space, on the theory that the effects
         of these states have already been included by the variational
         optimization of the reference.
        }
\label{fig:modified_h0}
\end{figure}

In our previous $N^7$-scaling version of the theory, we chose the zeroth
order Hamiltonian to have the following form.
\begin{align}
\label{eqn:old_H0}
\hat{H}_0 % ^{\mathrm{(old)}} 
          =   \hat{R} ( \hat{F} - \hat{H} ) \hat{R}
            + \hat{P} \hat{H} \hat{P}
            + \hat{Q} \hat{F} \hat{Q}
\end{align}
Here, we will retain this form, but make some modifications in the
triples space to improve efficiency.
As before, we take
%account for the fact that we are no longer including the closed
%shell piece in $\left|\Psi_0\right>$ and in order to reduce the cost
%of the theory.
%In particular, the revised theory defines
$\hat{F}$ to be the Fock operator constructed from the one-body
density matrix of $\left|\Psi_0\right>$,
$\hat{R}=\left|\Psi_0\right>\left<\Psi_0\right|$ to be the projector
on to the zeroth order wave function, $\hat{P}$
to be the projector on to the span of the closed shell determinant
$\left|\Phi\right>$ and all singly excited determinants,
and $\hat{Q}=1-\hat{P}$.
The difference between the present theory and our previous approach is
that, in the present theory, we work in the TOP orbital basis and
modify the $\hat{Q} \hat{F} \hat{Q}$ term so that it is diagonal
in some parts of the triples space.
To see how, let us first organize the triply excited determinants
into three groups:  the N-triples whose six indices $i$,$j$,$k$,$a$,$b$,$c$
do not contain any TOPs, the L-triples that contain at least one
TOP whose singular value from Eq.\ (\ref{eqn:psi0_tp}) is large
(above a threshold $\eta$), and the S-triples that contain at least
one TOP but whose TOPs all have small singular values (below $\eta$).
With the triples organized into these three groups, we make the modification
\begin{align}
    \hat{Q} \hat{F} \hat{Q} \hspace{1mm} \rightarrow \hspace{0mm}
    & \hspace{2mm}
      \big( \hat{Q}_{\mathrm{D}} + \hat{Q}_{\mathrm{L}} \big)
      \hspace{0.5mm} \hat{F} \hspace{0.5mm}
      \big( \hat{Q}_{\mathrm{D}} + \hat{Q}_{\mathrm{L}} \big)
    \notag \\
    & \hspace{5mm} + \hspace{2mm}
      \big( \hat{Q}_{\mathrm{S}} + \hat{Q}_{\mathrm{N}} \big)
      \hspace{0.5mm} \hat{F}^{\mathrm{(diag)}} \hspace{0.5mm}
      \big( \hat{Q}_{\mathrm{S}} + \hat{Q}_{\mathrm{N}} \big)
    \label{eqn:qfq_replacement}
\end{align}
in which $\hat{F}^{\mathrm{(diag)}}$ is the Fock operator with its
off-diagonal terms set to zero and
$\hat{Q}_{\mathrm{D}}$,
$\hat{Q}_{\mathrm{L}}$,
$\hat{Q}_{\mathrm{S}}$, and
$\hat{Q}_{\mathrm{N}}$
project on to the doubles, the L-triples, the S-triples, and the
N-triples, respectively.
As shown in Figure \ref{fig:modified_h0}, 
we are making $\hat{H}_0$ diagonal for the
presumably less important S-triples and N-triples, whereas our
previous approach left it off-diagonal for all triples.

How much efficiency is gained by this approach depends on how one
chooses to divide the TOP-containing triples between the large
and small L-triples and S-triples spaces.
In the $\eta=0$ extreme, in which all the TOP-containing triples
are placed in the L-triples space, the cost of forming the
right-hand-side of and solving the usual Rayleigh-Shr{\"o}dinger
linear equation for $\left|\Psi_1\right>$ grows as $N^6$.
The reduction from the $N^7$ scaling of our previous approach
comes from eliminating the N-triples, which as discussed
in Section \ref{sec:first_order_wfn} have no
Eq.\ (\ref{eqn:triples_element}) matrix elements
and thus can only contribute to the theory by coupling
through $\hat{H}_0$ to other parts of $\left|\Psi_1\right>$,
which is prevented by our modification in Eq.\ (\ref{eqn:qfq_replacement}).
In the other extreme, when only triples that contain the TOP with the
largest singular value are placed in the L-triples space and all other
TOP-containing triples are placed in the S-triples space,
the cost of setting up and solving the linear equation
for $\left|\Psi_1\right>$ grows as only $N^5$.
Note that, as we explain in Section \ref{sec:cost_scaling}, we have
explicitly verified these scalings (the lower of which is actually
$N_o^2N_v^3$) by log-log regressions on the floating-point operation
counts of each individual term entering in to the setup and
iterative solution of the linear equation.

To put these extremes in to perspective, we note that for a
size-intensive excitation, by which we mean one whose spatial extent
does not grow indefinitely as the system is enlarged,
the number of non-zero singular values in Eq.\ (\ref{eqn:psi0_tp})
will be constant with system size in the large system limit.
This implies that for size-intensive excitations, setting $\eta$ to
a small but non-zero threshold will result in both a size-intensive
excitation energy (a property we verify explicitly below)
and $N^5$ scaling.
Note that this efficiency gain is \textit{not} due to assuming
anything about the locality of electron correlation
(which if exploited as in some ground state methods \cite{Neese2013PNOCC}
could perhaps further lower the method's scaling)
but instead comes from the natural tendency of molecular excitations
to be localized.
Of course, in practice, the length scale needed to see this benefit
may be much larger than the simulation in a particular system,
so let us make a more concrete statement about the scaling.
If one limits $\left|\Psi_0\right>$ to have only $N_{\mathrm{TOP}}$
nonzero singular values regardless of the system size, then the
method has an $N^5$ scaling.
The obvious practical case where this approach should be useful is for
excitations that are dominated by a single configuration state function (CSF),
and thus for which only one singular value is nonzero anyways.

\subsection{Automated Implementation}
\label{sec:auto_implementation}

For the construction and solution of the linear equation for
$\left|\Psi_1\right>$, we have written a
simple for-loop generator.
The approach is to start with a symbolic representation of
the for-loops belonging to each orbital index, and then
to use Wick's theorem to derive the different contraction
schemes that connect indices and thus eliminate for-loops via
the resulting Kronecker delta functions.
This entire process is automated and includes the detailed
logic needed to a) identify which triples reside in the L-triples
space and thus must be included in the iterative solution
of the linear equation (triples in the S and N spaces are
not part of the iterative solver, as their part of the
linear equation is diagonal and can be inverted directly)
and b) avoid double counting redundant terms, such as
$T_{ijk}^{abc}$ and $T_{jik}^{bac}$.
Of course, the result is a code build of ``dumb'' loops, which
will not be cache-optimal, but does provide us with a correct
reference implementation to start from.
Further, it allows us to automatically implement careful operation
counting, such that each contraction can have its cost scaling
analyzed independently.
Having thus identified the most expensive term
(which turns out to be a contraction
%involved in the mixed-spin-L-triples
%to mixed-spin-L-triples mapping across the Fock operator
between the mixed-spin-L-triples and the Fock operator)
we have verified that by hand-coding this term in terms of dense
linear algebra, the cost can be reduced by more than an order of magnitude.
In future, we will work to convert all other contractions whose cost
is not trivial into dense linear algebra.
In the present study, however, our focus is not on a production-level code,
but instead on completing a detailed analysis of the cost-scaling as well as
the accuracy of the new $N^5$ approach.

\begin{figure}[t]
\centering
\includegraphics[width=8.0cm,angle=0,scale=1.0]{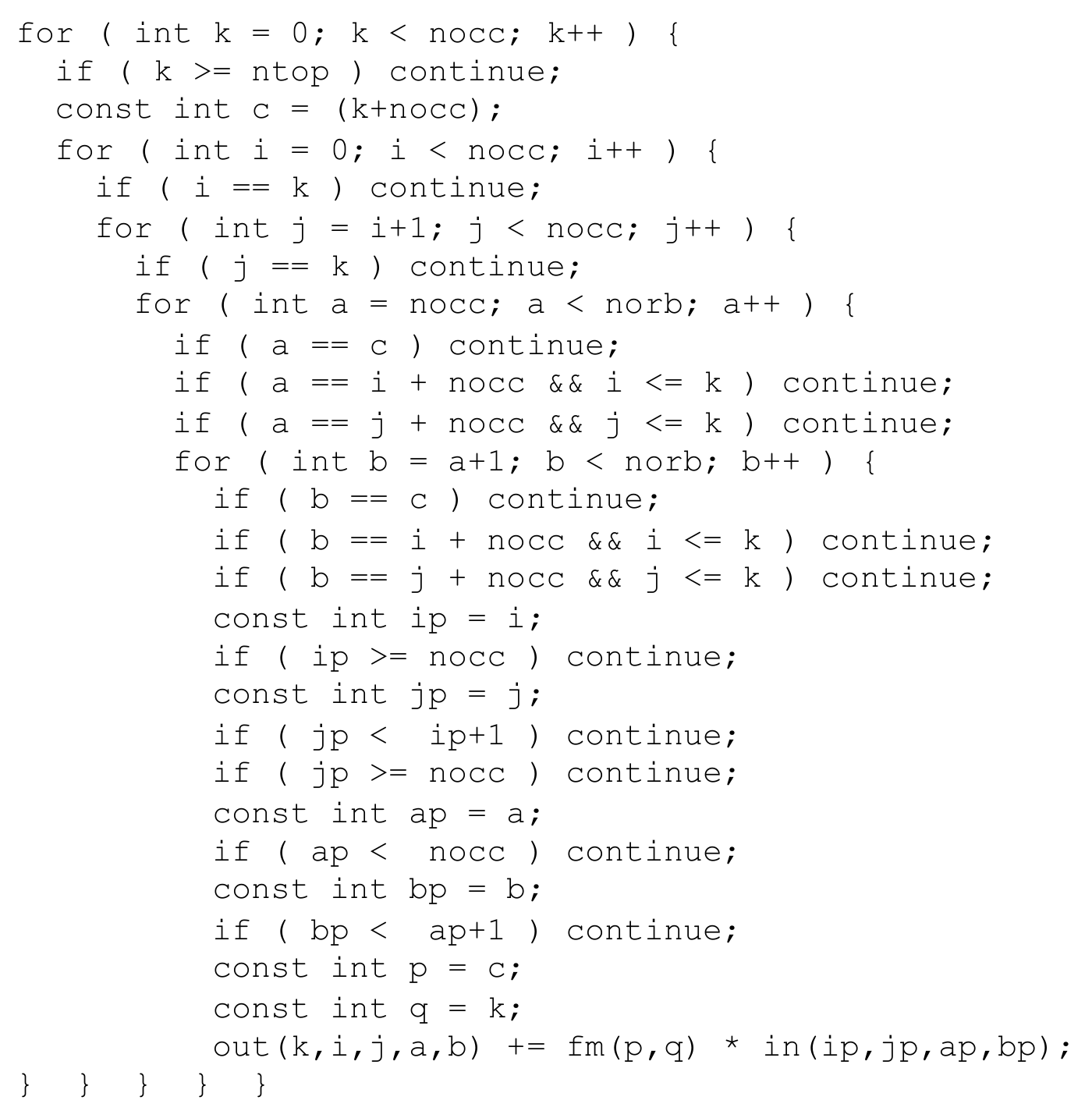}
\caption{The generated code for the
$\delta_{a a'} \delta_{b b'} \delta_{i i'} \delta_{j j'} \delta_{k q} \delta_{c p}$
         contraction resulting from Eq.\ (\ref{eqn:example_contraction}).
         Note the second line, where the scaling is explicitly
         reduced if the number $N_{\mathrm{TOP}}$ of TOPs that are considered
         large does not grow with system size.
        }
\label{fig:auto_loops}
\end{figure}

For an example of how the code generation works, consider how
the Fock operator might map double excitation coefficients to
L-triples excitation coefficients in the case where all orbitals
are spin up.
%  /    kt     a     b   |             |    ap    bp  \
%  |                     |   p^+    q  |              |  f_{p,q}  t_{ip, jp, ap, bp}
%  \     k     i     j   |             |    ip    jp  /
The corresponding tensor contractions come from the different
ways of contracting the indices in
\begin{align}
%Z_{kij}^{cab} & =
\sum_{pq} \hspace{1mm} \sum_{i'j'a'b'}
F_{pq} T_{i'j'}^{a'b'}
%\notag \\
%& \times
  \left<\Phi\right| \hat{\tau}^+_k \hat{\tau}^+_i \hat{\tau}^+_j
                    \hat{w}_b \hat{w}_a \hat{\sigma}_c
                    \hat{a}^+_p \hat{a}_q
                    \hat{w}^+_{a'} \hat{w}^+_{b'} \hat{\tau}_{j'} \hat{\tau}_{i'}
\left|\Phi\right>,
\label{eqn:example_contraction}
\end{align}
where the L-triple's indices are $i$,$j$,$k$,$a$,$b$,$c$ and
we assume, without loss of generality,
that $c$ and $k$ form a TOP such that
$c=k+N_o$ (at least one TOP must be
present as this is an L-triple).
The automatically generated code for one of the contractions
resulting from Eq.\ (\ref{eqn:example_contraction})
is seen in Figure \ref{fig:auto_loops},
where we see explicitly in the second line of code the simplification and
lower scaling that comes if we fix the number $N_{\mathrm{TOP}}$ of
large TOPs.
This particular term has $N_o^2N_v^2$ scaling if the number of
large-singular-value TOPs is fixed, or $N_o^3N_v^2$ if it grows
with system size (e.g.\ if all TOPs are considered large).
Across all the different pieces needed to construct the linear
equation's right-hand-side and to operate by $\hat{H}_0$,
the automated generator found 185 contractions with
non-zero contributions.
When $N_{\mathrm{TOP}}$ is set to one, only 18 of the contractions
involving triples showed fifth-order cost-scaling, and only 9 of
those showed the most expensive $N_o^2N_v^3$ scaling, suggesting
that converting the worst terms to hand-coded dense linear algebra
(i.e.\ BLAS) should be feasible in future work.
See section \ref{sec:cost_scaling} below for a more detailed
cost scaling analysis.

\section{Results}
\label{sec:results}

\subsection{Computational Details}
\label{sec:comp_details}

%% We used the iterative conjugate-gradient algorithm to solve
%% the linear equation for $\left|\Psi_1\right>$.
%% 
%% There reference $\delta$-CR-EOM-CC(2,3) \cite{piecuch:delta_eom_ccsd}
%% calculations were performed with GAMESS.\cite{gamess1}
%% CASPT2 calculations were performed with Molpro.
%% \cite{werner2012molpro,werner1996caspt3,celani2000multireference}

For ESMP2, we used the iterative conjugate-gradient
algorithm to solve the linear equation for $\left|\Psi_1\right>$.
The EOM-CCSD and $\delta$-CR-EOM-CC(2,3) \cite{piecuch:delta_eom_ccsd, piecuch2002crccsd, piecuch2009lcreomcc, Kowalski2004escc}
calculations were performed with GAMESS,\cite{gamess1, gamess2}
whereas CIS and TDDFT calculations were performed
with QChem. \cite{qchem}
CASPT2 calculations for the ring excitations were performed with Molpro.
\cite{werner2012molpro,werner1996caspt3,celani2000multireference}
Note that, although some CASPT2 calculations relied on state-averaged
CASSCF reference functions, the CASPT2 calculations
themselves were single-state.
In the pyrrole molecule, CASSCF was performed with
a (10o,6e) active space, in which an equal-weight
4-state state-average was employed, with 2, 1, and 1
states from the A$_1$, A$_2$, and B$_2$ representations,
respectively.
Note that pyrrole's $2^1$A$_1$ excitation
was not stable in CASPT2 without a level shift, and
so in this molecule a level shift of 0.2 E$_h$ was
used for all states.
In the rest of the ring excitations, CASPT2 was stable
without a level shift, and so no shift was used in
other molecules.
For pyridine, CASSCF employed an (8o,10e) active space
and four separate state averaging calculations, one in each
representation of its C$_{2v}$ point group.
No level shift was necessary for stability in pyridine, but
each of these four calculations used an equal-weight
3-state state average, which was necessitated by the
fact that at the CASSCF level the $1^1$B$_2$ and
$2^1$B$_2$ states come first and third in the energy ordering
of the ${}^1$B$_2$ states.
For benzene, CASSCF employed a (6o,6e) active space
for an equal-weight 6-state state-average with two states
each in the A$_g$, B$_{1u}$, and B$_{2u}$ representations
(the computational point group was D$_{2h}$).
Finally, for pyrimidine, CASSCF employed an (8o,10e) active space.
As all the states investigated in pyrimidine are ground
states within their own symmetries,
state averaging was not used in this case.

\subsection{Cost Scaling Analysis}
\label{sec:cost_scaling}

Before looking at the energetic accuracy of reduced-scaling
ESMP2, let us first inspect how the different
components scale with the number of occupied and virtual orbitals.
Using our automatic code generator, we have inserted operation
counting into all of the terms, allowing for a
contraction-by-contraction scaling analysis.
For each contraction, we measured occupied scaling by
fixing the number of virtual orbitals at 100 and varying the
number of occupied orbitals between 30 and 50, after which
we perform a log-log linear regression on each contraction's
operation count.
Similarly, for scaling with virtual orbitals, we have fixed
the number of occupied orbitals at 30 and varied the number
of virtuals between 50 and 100, again feeding the information
into log-log linear regressions.
We present two sets of scaling data, representing the
two cost extremes that one can get from the new approach.
First, Table \ref{tab:scaling_1nto} shows the scaling data
for ESMP2-TOP(1), in which only one TOP is considered large.
Second, Table \ref{tab:scaling_allnto} shows the scaling data
for ESMP2-TOP(all), in which all TOPs are considered large
and the S-triples space is thus empty.

This detailed scaling analysis reveals that the worst-scaling
terms all reside in the linear transformation part of solving
the PT2 linear equation, which is to say evaluating the action
of $H_0$ on a vector in the first-order interacting space as
required in each iteration of the conjugate gradient algorithm
we use to solve the linear equation.
These terms are fifth and sixth order in the system size
for ESMP2-TOP(1) and ESMP2-TOP(all), respectively, showing
that it is indeed possible to improve over the seventh order
scaling of the original formulation of ESMP2.
Of course, prefactors matter,
and the fact that ESMP2-TOP(1) carries 13 terms with
$N_o^2N_v^3$ scaling means that, for small systems, it will
almost certainly be slower than EOM-CCSD despite its
lower scaling, as EOM-CCSD has a smaller number of
$N^5$ and $N^6$ terms.
The scaling does guarantee, though, that ESMP2-TOP(1) will
be faster in larger systems.
We now turn our attention to
% two important followup questions,
% which is whether the number of large TOPs can be expected to
% grow with system size, and then whether limiting the number
% of large TOPs results in energetic accuracy similar
% to the original seventh order formulation.
the question of whether energetic accuracy is maintained
when we aggressively limit the number of TOPs that are
considered large.

\begin{table}[t]
\caption{
  \label{tab:scaling_1nto}
  Scaling data for ESMP2-TOP(1), in which only one TOP is treated
  as large and which is thus most appropriate when the reference
  is dominated by a single CSF.
  For the different parts of the PT2 linear equation's right-hand side (RHS)
  and linear transformation,
  we report how many of that part's contractions
  fall in to the different asymptotic scaling categories.
  %RHS refers to the right-hand-side expression in the PT2 linear equation
  %while $H_0$ A$\leftrightarrow$B refers to the operations involved
  %in the $H_0$ linear mapping between amplitudes of types A and B.
  %Note that in cases where the log-log scaling regression exponents for
  %occupied or virtual orbitals
  %differed from integers by more than 0.25, we rounded up the virtual
  %exponent and rounded down the occupied exponent to present
  %a conservative estimate of the asymptotic scaling.
  %For example, $N_o^{1.4}N_v^{1.5}$ is converted to $N_oN_v^2$.
}
\begin{tabular}{l c c c c c}
\hline\hline
\multicolumn{1}{l}{ \hspace{0mm} \rule{0pt}{3.8mm} } &
\hspace{2mm} $N^3$ &
\hspace{2mm} $N_o^3N_v$ &
\hspace{2mm} $N_o^2N_v^2$ &
\hspace{2mm} $N_o^3N_v^2$ &
\hspace{2mm} $N_o^2N_v^3$
\\[2pt]
\hline
\hspace{0.0mm} $T_{ij}^{ab}$ \rule{0pt}{3.8mm}(RHS)
     &  4 & 12 &  6 &  0 &  0 \\[3pt]
\hspace{0.0mm} $S_{ij}^{ab}$ (RHS)
     &  2 &  6 &  3 &  0 &  0 \\[3pt]
\hspace{0.0mm} $T_{ijk}^{abc}$ (RHS)
     & 16 &  0 &  4 &  0 &  0 \\[3pt]
\hspace{0.0mm} $S_{ijk}^{abc}$ (RHS)
     &  3 &  0 &  5 &  0 &  0 \\[3pt]
\hspace{0.0mm} $T_{ij}^{ab}\leftrightarrow T_{ij}^{ab}$
     &  0 &  0 &  1 &  2 &  2 \\[3pt]
\hspace{0.0mm} $S_{ij}^{ab}\leftrightarrow S_{ij}^{ab}$
     &  0 &  0 &  1 &  2 &  2 \\[3pt]
\hspace{0.0mm} $T_{ijk}^{abc}\leftrightarrow T_{ijk}^{abc}$
     &  0 &  0 &  3 &  4 &  4 \\[3pt]
\hspace{0.0mm} $S_{ijk}^{abc}\leftrightarrow S_{ijk}^{abc}$
     &  0 &  0 &  2 &  5 &  5 \\[3pt]
\hspace{0.0mm} $T_{ijk}^{abc}\leftrightarrow T_{ij}^{ab}$
     &  0 &  0 & 18 &  0 &  0 \\[3pt]
\hspace{0.0mm} $S_{ijk}^{abc}\leftrightarrow S_{ij}^{ab}$
     &  0 &  0 &  8 &  0 &  0 \\[3pt]
\hspace{0.0mm} $S_{ijk}^{abc}\leftrightarrow T_{ij}^{ab}$
     &  0 &  0 &  2 &  0 &  0 \\[3pt]
\hline\hline
%\vspace{2mm}
\end{tabular}
\end{table}

\begin{table}[t]
\caption{
  \label{tab:scaling_allnto}
  Scaling data for ESMP2-TOP(all),
  in which all TOPs are treated as large.
  For the different parts of the PT2 linear equation's
  right-hand side (RHS) and linear transformation,
  we report how many of that part's contractions
  fall in to the different asymptotic scaling categories.
  We omit the RHS doubles terms and the doubles-only parts
  ($T_{ij}^{ab}\leftrightarrow T_{ij}^{ab}$ and
   $S_{ij}^{ab}\leftrightarrow S_{ij}^{ab}$)
  of the linear transformation, as their scaling is the same
  as in Table \ref{tab:scaling_1nto}.
  Note that in cases where the log-log scaling regression exponents for
  occupied or virtual orbitals were significantly fractional
  (i.e.\ differed from integers by more than 0.3)
  we took the conservative approach of transferring enough fractional
  exponent from occupied to virtual in order to move the virtual
  exponent up to the next integer, and then rounded what remained
  of the fractional occupied exponent up or down if it was above
  or below 0.3.
  For example, $N_o^{2.6}N_v^{2.5}$ is converted to $N_o^2N_v^3$,
  while $N_o^{2.99}N_v^{1.35}$ is converted to $N_o^3N_v^2$.
  Note that, although this conservative rounding may slightly
  rearrange the contractions among the $N^4$ and $N^5$
  categories, we have explicitly verified (by inspecting the code)
  that each of the $N^6$ contractions has the asymptotic scaling
  reported here.
}
\begin{tabular}{l c c c c c c c}
\hline\hline
\multicolumn{1}{l}{ \hspace{0mm} \rule{0pt}{3.8mm} } &
\hspace{0.2mm} $N^4$ &
\hspace{0.9mm} $N_o^5$ &
\hspace{0.9mm} $N_o^4N_v$ &
\hspace{0.9mm} $N_o^3N_v^2$ &
\hspace{0.9mm} $N_o^2N_v^3$ &
\hspace{0.9mm} $N_o^4N_v^2$ &
\hspace{0.9mm} $N_o^3N_v^3$
\\[2pt]
\hline
\hspace{0.0mm} $T_{ijk}^{abc}$ \rule{0pt}{3.8mm}(RHS)
     & 16 &  0 &  0 &  4 &  0 &  0 &  0 \\[3pt]
\hspace{0.0mm} $S_{ijk}^{abc}$ (RHS)
     &  3 &  0 &  0 &  5 &  0 &  0 &  0 \\[3pt]
\hspace{0.0mm} $T_{ijk}^{abc}\leftrightarrow T_{ijk}^{abc}$
     &  0 & 25 &  8 & 13 & 10 &  7 &  4 \\[3pt]
\hspace{0.0mm} $S_{ijk}^{abc}\leftrightarrow S_{ijk}^{abc}$
     &  0 &  2 &  5 &  6 &  1 &  5 &  5 \\[3pt]
\hspace{0.0mm} $T_{ijk}^{abc}\leftrightarrow T_{ij}^{ab}$
     &  0 & 10 & 10 & 20 & 10 &  0 &  0 \\[3pt]
\hspace{0.0mm} $S_{ijk}^{abc}\leftrightarrow S_{ij}^{ab}$
     &  0 &  0 &  0 &  8 &  0 &  0 &  0 \\[3pt]
\hspace{0.0mm} $S_{ijk}^{abc}\leftrightarrow T_{ij}^{ab}$
     &  0 &  0 &  4 &  2 &  2 &  0 &  0 \\[3pt]
\hline\hline
%\vspace{2mm}
\end{tabular}
\end{table}

\begin{table*}[t]
\caption{
  \label{tab:small_molec}
  For the lowest singlet excitations in several small molecules,
  as well as for two simple CT excitations, we report the
  reference $\delta$-CR-EOM-CC(2,3) excitation energy in eV,
  as well as other methods' errors relative to the reference.
  All calculations are in the cc-pVDZ basis.
  Only the dominant TOPs were considered large in ESMP2,
  meaning one TOP in all cases except N$_2$, where by symmetry
  there are two dominant TOPs with equal weights.
  % $H_0$ was diagonal in the space of triples that do not
  % contain the dominant TOP and that the method's scaling
  % with system size is fifth-order.
  Below each method, we report the canonical cost scaling
  with respect to system size.
  At bottom, we report mean and maximum absolute (i.e.\ unsigned)
  deviations from the reference both with and without the CT
  systems included, as well as the number of deviations
  larger than 0.3 eV.
}
\begin{tabular}{l c r r r r r}
\hline\hline
\rule{0pt}{3.8mm} &
\hspace{1mm} $\delta$-CR-EOM-CC(2,3) &
\hspace{2mm} CIS &
\hspace{2mm} TDDFT/B3LYP &
\hspace{2mm} TDDFT/$\omega$B97X &
\hspace{2mm} EOM-CCSD &
%\hspace{2mm} ESMF & &
\hspace{2mm} ESMP2 \\
 & \multicolumn{1}{c}{\hspace{0mm} $O(N^7)$}
 & \multicolumn{1}{r}{\hspace{1mm} $O(N^4)$}
 & \multicolumn{1}{r}{\hspace{0mm} $O(N^4)$}
 & \multicolumn{1}{r}{\hspace{0mm} $O(N^4)$}
 & \multicolumn{1}{r}{\hspace{0mm} $O(N^6)$}
 & \multicolumn{1}{r}{\hspace{0mm} $O(N^5)$}
\\[2pt]
\hline
\rule{0pt}{3.8mm}Acetaldehyde $1^1A"$
                       &  4.36 &  0.71 &  0.09 &  0.14 &  0.21 &  0.16 \\
Ammonia  $2^1A_1$     &  7.57 &  0.95 & -0.52 & -0.07 &  0.05 &  0.01 \\
Carbon Monoxide  $1^1\Pi $      &  8.76 &  0.61 &  0.16 &  0.31 &  0.30 & -0.09 \\ % not sure about the character
Cyclopropene  $2^1B2$         &  7.97 &  0.57 & -0.83 & -0.33 & -0.08 & -0.07 \\
Diazomethane  $1^1A_2$         &  3.01 &  0.38 &  0.05 &  0.09 &  0.45 & -0.00 \\
Dinitrogen  $1\Pi_g$           & 10.36 & -1.31 & -0.03 &  0.00 &  0.44 &  0.09 \\ % not sure about the character
Ethylene  $1^1B_3$             &  8.80 & -0.25 &  0.11 &  0.10 &  0.19 & -0.30 \\
Formaldehyde  $1^1A_2$         &  4.08 &  0.63 &  0.07 &  0.10 &  0.19 &  0.15 \\
Formamide  $2^1A"$            &  5.86 &  0.88 &  0.04 &  0.11 &  0.21 &  0.15 \\
Hydrogen Sulfide  $2^1B_2$      &  7.05 &  0.58 & -0.27 &  0.20 &  0.11 & -0.07 \\
Ketene  $1^1A_2$              &  3.78 &  0.70 &  0.22 &  0.31 &  0.36 & -0.01 \\
Methanimine  $1^1A"$       &  5.35 &  0.66 &  0.00 &  0.11 &  0.22 & -0.00 \\
Nitrosomethane  $1^1A"$       &  1.85 &  0.27 &  0.13 &  0.12 &  0.25 &  0.17 \\
Streptocyanine Cation $1^1B_2$ &  7.53 &  1.55 &  1.08 &  1.07 &  0.28 & -0.40 \\
Thiofromaldyhyde $1^1A_2$      &  2.18 &  0.58 &  0.13 &  0.17 &  0.24 & -0.08 \\
Water  $1^1B_2$                &  8.30 &  1.02 & -0.57 & -0.22 & -0.01 &  0.06 \\[2pt]
%\rule[0.91ex]{0pt}{0pt}
\hline
\rule{0pt}{3.8mm}Ammonia $\to$ Diflourine  $2^1A_1$
                           &  9.27 & 2.38 & -6.91 & -2.69 & 0.51 & -0.26 \\
Dinitrogen $\to$ Methylene $1^1B_2$ & 15.49 & 1.66 & -6.58 & -1.79 & 0.06 &  0.15 \\[2pt]
\hline
\multicolumn{2}{l}{\rule{0pt}{3.8mm}Mean Abs. Dev. (with CT)}
                                   &  0.87 & 0.99 &  0.44 & 0.23 &  0.12 \\
\multicolumn{2}{l}{Max Abs. Dev. (with CT)}
                                   &  2.38 & 6.91 &  2.69 & 0.51 &  0.40 \\[2pt]
\hline
\multicolumn{2}{l}{\rule{0pt}{3.8mm}Mean Abs. Dev. (without CT)}
                                   &  0.73 & 0.27 &  0.22 & 0.22 &  0.11 \\
\multicolumn{2}{l}{Max Abs. Dev. (without CT)}
                                   &  1.55 & 1.08 &  1.07 & 0.45 &  0.40 \\[2pt]
\hline
\multicolumn{2}{l}{\rule{0pt}{3.8mm}Deviations above 0.3 eV}
                                   &  16~~ &   6~~ &   6~~ &  4~~ & 1~~ \\[2pt]
%\rule[1.0ex]{0pt}{0pt} \\
\hline\hline
\vspace{2mm}
\end{tabular}
\end{table*}

\subsection{Small Molecule Testing}
\label{sec:small_molec}

Let us begin by testing the fifth order method on the same set of
small molecules and two charge transfer (CT) examples that were
studied recently with the original seventh order incarnation of
ESMP2.
To make the comparison direct, we use the same cc-pVDZ basis
and the same molecular geometries as in the previous study.
\cite{Shea2020GVP}
Here, we restrict the L-triples space as much as possible
by treating only the dominant TOP (or, in the case of N$_2$,
the pair of equal-weight dominant TOPs) as large,
relegating triples that do not contain the dominant TOP
to the S-triples space with its diagonally-approximated zeroth
order Hamiltonian.
In Table \ref{tab:small_molec}, we see that for this set
of small molecules, this $N_o^2N_v^3$-scaling variant
of ESMP2 is, like its seventh order predecessor,
competitive in accuracy with EOM-CCSD.
Thus, by working with excited-state-specific orbitals
from the ESMF reference and an excited-state-specific
correlation treatment from ESMP2, it is possible,
at least in these test systems, to achieve EOM-CCSD
accuracies with a method that scales as the fifth order
of the system size.
As with the original formulation of ESMP2, we find it
especially encouraging that the method is equally accurate
for CT and non-CT states, as practical uses of CT
in biological and energy-related chemistry often
involve large system sizes where lower-scaling methods
are essential.

Interestingly, the results here are barely changed compared
to the results from the previous seventh order method,
which displayed mean absolute errors of 0.13 eV and 0.12 eV
for the full set and the non-CT subset, respectively.
\cite{Shea2020GVP}
This finding suggests that the basic idea here is sound:
using a diagonal approximation to $H_0$ in the space of
less-important triples does not have a significant effect
on the accuracy.
Note that we have tested whether having any off-diagonal $H_0$
character in the triples manifold is necessary by testing
what happens if no TOPs are treated as large.
In that case, we find that accuracy suffers significantly,
suggesting that, for the triples that connect directly
via the coulomb operator to the large parts of the zeroth
order reference, the fact that the Fock operator is
not diagonal is significant.
Thus, it appears that we get away with the reduction
in scaling not because the off-diagonal parts of the
Fock operator are unimportant, but because their
effects are small for the triples that
do not connect to the reference or that only connect
to small components (TOPs with small weights)
of the reference.

\begin{table*}[t]
\caption{
  \label{tab:rings}
  Excitation energies (eV) for small ring systems in the cc-pVDZ basis.
  For the $\delta$-CR-EOM-CC(2,3) reference, we report the excitation energy,
  while for other methods we report deviations from the reference.
  ESMP2 treated TOPs with singular values above 0.1 as large,
  which led to two or fewer large TOPs in all states included here.
  A diagonal $H_0$ was used in the space of triples that do not
  contain any large TOPs.
  At the bottom, we report mean and maximum absolute
  deviations from the reference, the number of these deviations that were
  larger than 0.3 eV, and the mean absolute deviation from CASPT2.
}
\begin{tabular}{l c r r r r r@{}l r}
\hline\hline
\rule{0pt}{3.8mm} State &
\hspace{1mm} $\delta$-CR-EOM-CC(2,3) &
\hspace{2mm} CIS &
\hspace{2mm} TDDFT/B3LYP &
\hspace{2mm} TDDFT/$\omega$B97X &
\hspace{2mm} EOM-CCSD &
\hspace{2mm} CASPT2 & &
\hspace{2mm} ESMP2 \\[2pt]
\hline
\rule{0pt}{3.8mm}Pyrrole  $2^1$A$_1$  
                       & 6.15 &  1.60 &  0.45 &  0.84 & 0.51 & -0.18 &${}^{*}$ & -0.90 \\
Pyrrole  $1^1$A$_2$    & 6.39 &  0.86 & -0.48 &  0.70 & 0.36 &  0.10 &         &  0.09 \\
Pyrrole  $1^1$B$_2$    & 6.56 &  0.37 &  0.01 &  0.10 & 0.47 &  0.38 &         & -0.21 \\
Pyridine $1^1$B$_1$    & 4.84 &  1.33 & -0.01 &  0.36 & 0.44 &  0.10 &         &  0.11 \\
Pyridine $1^1$B$_2$    & 4.76 &  1.44 &  0.75 &  0.83 & 0.52 &  0.07 &         & -0.25 \\
Pyridine $2^1$B$_2$    & 6.51 &  2.05 &  0.86 &  1.05 & 0.45 &  0.29 &         &  0.11 \\
Pyridine $1^1$A$_2$    & 5.26 &  2.19 & -0.16 &  0.33 & 0.44 & -0.05 &         & -0.05 \\
Benzene  $1^1$B$_{2u}$ & 4.69 &  1.33 &  0.72 &  0.83 & 0.50 &  0.06 &         & -0.71 \\
Benzene  $1^1$B$_{1u}$ & 6.35 & -0.08 & -0.21 & -0.06 & 0.42 & -0.35 &         & -0.26 \\
Benzene  $2^1$B$_{1u}$ & 7.33 &  0.94 & -0.14 & -0.01 & 0.43 & -0.69 &         & -0.82 \\
Pyrimidine $1^1$B$_1$  & 4.50 &  1.40 & -0.21 &  0.18 & 0.22 & -0.32 &         & -0.82 \\
Pyrimidine $1^1$B$_2$  & 5.23 &  1.28 &  0.52 &  0.61 & 0.28 & -0.22 &         & -0.28 \\[2pt]
%Cl$^-$-H$_2$O & 4.7195 & 4.7195 \\
%NH$_3$-F$_2$ & 4.5367 & 4.5367 \\
\hline
\multicolumn{2}{c}{\rule{0pt}{3.8mm}Mean Abs. Dev. (MAD)}
                              &  1.24 &  0.38 &  0.49 & 0.42 &  0.23 &         &  0.38 \\
\multicolumn{2}{c}{Max Abs. Dev.}
                              &  2.19 &  0.86 &  1.05 & 0.52 &  0.69 &         &  0.90 \\
\multicolumn{2}{c}{Deviations above 0.3 eV}
                              &  11~~ &   6~~ &   8~~ & 10~~ &   4~~ &         &   4~~ \\
\multicolumn{2}{c}{MAD vs CASPT2}
                              &  1.30 &  0.44 &  0.59 & 0.49 &  0.00 &         &  0.28 \\[2pt]
\hline\hline
\multicolumn{9}{l}{
${}^{*}$Level shift was necessary for convergence.  See text.\rule{0pt}{3.8mm}
} \\
%\vspace{2mm}
\end{tabular}
\end{table*}

\begin{table*}[t]
\caption{
  \label{tab:rings_TZ}
  Excitation energies (eV) for small ring systems in the cc-pVTZ basis.
  For the $\delta$-CR-EOM-CC(2,3) reference, we report the excitation energy,
  while for other methods we report deviations from the reference.
  ESMP2 treated TOPs with singular values above 0.1 as large,
  which led to two or fewer large TOPs in all states included here.
  A diagonal $H_0$ was used in the space of triples that do not
  contain any large TOPs.
  At the bottom, we report mean and maximum absolute
  deviations from the reference and the number of these deviations that were
  larger than 0.3 eV.
}
\begin{tabular}{l c r r}
\hline\hline
\rule{0pt}{3.8mm} State &
\hspace{1mm} $\delta$-CR-EOM-CC(2,3) &
\hspace{2mm} EOM-CCSD &
\hspace{2mm} ESMP2 \\[2pt]
\hline
\rule{0pt}{3.8mm}Pyrrole  $2^1$A$_1$  
                       & 5.95 & 0.59 & -0.89\\
Pyrrole  $1^1$A$_2$    & 5.93 & 0.41 &  0.13 \\
Pyrrole  $1^1$B$_2$    & 6.25 & 0.15 & -0.18 \\
Pyridine $1^1$B$_1$    & 4.69 & 0.52 &  0.15 \\
Pyridine $1^1$B$_2$    & 6.22 & 0.51 &  0.12 \\
Pyridine $2^1$B$_2$    & 4.61 & 0.60 & -0.21 \\
Pyridine $1^1$A$_2$    & 5.14 & 0.51 & -0.03 \\
Benzene  $1^1$B$_{2u}$ & 4.55 & 0.58 & -0.66 \\
Benzene  $1^1$B$_{1u}$ & 7.01 & 0.51 & -0.76 \\
Benzene  $2^1$B$_{1u}$ & 6.06 & 0.48 & -0.23 \\
Pyrimidine $1^1$B$_1$  & 4.14 & 0.54 & -0.60 \\
Pyrimidine $1^1$B$_2$  & 4.82 & 0.62 & -0.23 \\[2pt]
%Cl$^-$-H$_2$O & 4.7195 & 4.7195 \\
%NH$_3$-F$_2$ & 4.5367 & 4.5367 \\
\hline
\multicolumn{2}{c}{\rule{0pt}{3.8mm}Mean Abs. Dev. (MAD)}
                              &  0.51 & 0.35 \\
\multicolumn{2}{c}{Max Abs. Dev.}
                              &  0.62 & 0.89 \\
\multicolumn{2}{c}{Deviations above 0.3 eV}
                              & 11~~ &  4~~ \\[2pt]
\hline\hline
%\vspace{2mm}
\end{tabular}
\end{table*}

\subsection{Ring Excitations}
\label{sec:multi_csf_systems}

We now turn to a set of low-lying excitations in aromatic
ring systems, where it is common to see excited states
in which more than one TOP has a large weight.
For these systems
(whose geometries have been taken from the cc-pVDZ MP2
entries in the CCCBDB NIST database \cite{cccbdb})
we have defined large TOPs as those whose singular
values are above 0.1, resulting in two or fewer
large TOPs in each excitation and thus a method that
remains at the fifth-order end of the continuum
between ESMP2-TOP(1) and ESMP2-TOP(all).
Unlike the small molecules of the previous section,
some of these ring excitations have at least a modest
(although not dominant) degree of doubly excited character.
As CASPT2 is often used to address double excitations,
we have also included a comparison against its results
in the table, although we stress that
$\delta$-CR-EOM-CC(2,3) is the better
reference in these states thanks to its ability to handle
double excitations and its higher-order treatment
of electron correlation.
This comparison makes clear that, at least on average,
ESMP2 is more similar to CASPT2 than to 
$\delta$-CR-EOM-CC(2,3), which is perhaps not surprising
given the fact that ESMP2 and CASPT2 approach these states
via second order perturbation theory from
a qualitatively correct reference, making them
methodologically similar.
Of course, the fact that ESMP2 need not specify an
active space is a significant practical advantage.
%approach is not
%necessarily quite as trustworthy as a reference here.
%This in mind, we have also made comparisons to CASPT2
%results, although this is not a perfect reference either.
%While CASPT2 is far more robust in the presence of
%doubly excited character, its weak correlation treatment
%is only second order, as opposed to the third order
%treatment of $\delta$-CR-EOM-CC(2,3).

\begin{table}[h]
\caption{
   Results for a few Rydberg states in neon, formaldehyde, and benzene. Molecular geometries were taken from the NIST CCCBDB database.\cite{cccbdb} All values are reported in eV.
   \label{tab:rydberg}
}
\begin{center}
\begin{tabular}{l l c c c}
\hline\hline
\rule{0pt}{3.8mm} State                    & 
\hspace{0.0mm} Basis     \hspace{0.0mm} & 
\hspace{0.0mm} ESMF    \hspace{0.0mm} &
\hspace{0.0mm} N5-ESMP2    \hspace{0.0mm} &
\hspace{0.0mm} Error \hspace{0.0mm}\\
\hline
% State                         Basis                ESMF       N5-ESMP2   error
  \rule{0pt}{3.8mm}Neon (2s $\rightarrow$ 3p)   & cc-pVTZ           & 65.6781  & 64.6521  &  0.35$^a$ \\ 
  Formaldehyde 2$^1$A$_1$      & d-aug-cc-pVTZ     &  7.0967  &  8.3287  &  0.23$^b$ \\ 
  Formaldehyde 3$^1$A$_1$      & d-aug-cc-pVTZ     &  8.1856  &  9.3947  &  0.13$^b$ \\ 
  Benzene 1E$_{2g}$            & aug-ANO1$^c$      &  6.7758  &  7.4583  & -0.38$^c$ \\ 
  Benzene 2A$_{1g}$            & aug-ANO1$^c$      &  6.7619  &  7.4557  & -0.39$^c$ \\ 
  Benzene 1A$_{2g}$            & aug-ANO1$^c$      &  6.8013  &  7.4856  & -0.38$^c$ \\ 
\hline\hline
\multicolumn{5}{l}{\rule{0pt}{3.8mm}$^a$ Compared to EOM-CCSD in the same basis.\cite{Shea2018}} \\
\multicolumn{5}{l}{$^b$ Compared to the theoretical best estimate for these states.\cite{Loos2020MediumMountaineering}} \\
\multicolumn{5}{l}{$^c$ Compared to CCSD calculations in the same basis set.\cite{aromatic_benchmarking}}
\end{tabular}
\end{center}
\end{table}

Across the twelve ring excitations shown in Table
\ref{tab:rings}, we find that the differences
between EOM-CCSD and ESMP2 are more significant
than in the small-molecule excitations of the
last section.
While EOM-CCSD has a slightly higher mean absolute
deviation from $\delta$-CR-EOM-CC(2,3),
its deviations are more regular than those of ESMP2.
Indeed, in all twelve cases, EOM-CCSD predicts
excitation energies to be between 0.2 and 0.55 eV
higher than does $\delta$-CR-EOM-CC(2,3), whereas
the span of ESMP2's deviations is significantly
larger at just over an eV.
Table \ref{tab:rings_TZ} shows that a similar story
plays out for EOM-CCSD and ESMP2 in a triple-zeta
basis, reassuring us that these tendencies are
not specific to the double-zeta basis, on which
we now focus our attention.

Notably, while Table \ref{tab:rings}'s ESMP2 results
are within 0.3 eV of
$\delta$-CR-EOM-CC(2,3) for eight out of the twelve
states, in the other four states
---
pyrrole $2^1$A$_1$,
benzene $1^1$B$_{2u}$,
benzene $2^1$B$_{1u}$, and
pyrimidine $1^1$B$_1$ ---
its prediction is low by 0.7 eV or more.
Two of these are errors likely due to doubly excited
character, one an error related to intruder states issues,
and one is not necessarily much of an error at all.
Start with the $2^1$A$_1$ state of pyrrole, where CASPT2
displays intruder-state behavior and is not
stable without the application of a level shift.
Given that the zeroth order Hamiltonians are
similar, and that the CASSCF reference used by
CASPT2 should be a better starting point than ESMF,
ESMP2's difficulty in this state is likely related to
these intrude state difficulties.
In the $2^1$B$_{1u}$ state of benzene, on the other hand,
ESMP2 is energetically very similar to CASPT2,
which is known to be highly accurate for the low-lying
excitations of benzene,
\cite{Roos1995Benzene,Bartlett1997EOMCCSDBenzene,Loos2020MediumMountaineering}
and so this appears to be a case where ESMP2 is
reasonably accurate, at least if CASPT2 is used as the reference.
Indeed, the MAD of ESMP2 relative to CASPT2 across all twelve states is
significantly lower than its MAD relative to $\delta$-CR-EOM-CC(2,3),
which is perhaps not so surprising given that both
ESMP2 and CASPT2 are second-order perturbation theories
based on orbital-optimized reference functions
(although for CASPT2 the orbital optimization is state-averaged,
rather than state-specific).
However, the agreement is certainly not perfect, and the large
deviations between ESMP2 and $\delta$-CR-EOM-CC(2,3) in the
benzene $1^1$B$_{2u}$ and pyrimidine $1^1$B$_1$ states cannot be
explained by either similarity to CASPT2 or by intruder state
issues in CASPT2, which were not present.
The errors in these two states are likely due instead to
doubly excited character that the singly-excited ESMF reference
function cannot capture.
Indeed, the doubly excited fractions of the CASSCF wave functions
for benzene $1^1$B$_{2u}$ and pyrimidine $1^1$B$_1$
were 15\% and 8\%, respectfully.
It is interesting to note that, at least in these two cases,
this modest fraction of doubly excited character caused
less trouble for EOM-CCSD.
This raises the interesting question of whether, for cases
with modest amounts of doubly excited character, EOM-CCSD
is more robust than ESMP2, which seems like a question
worth studying more systematically in future work.

\subsection{Rydberg Excitations}
\label{sec:rydberg}

To check whether ESMP2 achieves a similar quality in Rydberg excitations,
we have tested it on relevant excitations in neon, formaldehyde, and benzene.
Although the comparison is less straightforward than those of the previous
sections due to a lack of a single high-level benchmark, comparisons to
literature values are shown in Table \ref{tab:rydberg}.
We find that the overall accuracy for ESMP2 is similar to that seen in the
ring systems above and that it makes a substantial correction
to the uncorrelated ESMF reference, which in most cases underestimates
these excitations (although interestingly not in neon).

\begin{table}[t]
\caption{
   Size intensivity test, in which we report the first singlet
   excitation energy in eV for a water molecule surrounded by
   a variable number of distant He atoms.
   Methods' asymptotic cost scalings are
   given in parentheses.
   \label{tab:size_test}
}
\begin{center}
\begin{tabular}{c c c c c c}
\hline\hline
\rule{0pt}{3.8mm}He                    & 
\hspace{0.0mm} ESMF     \hspace{0.0mm} & 
\hspace{0.0mm} ESMP2    \hspace{0.0mm} &
\hspace{0.0mm} ESMP2    \hspace{0.0mm} &
\hspace{0.0mm} EOM-CCSD \hspace{0.0mm} &
\hspace{0.0mm} CISD     \hspace{0.0mm} \\
atoms & (N$^4$) & (N$^5$) & (N$^7$) & (N$^6$) & (N$^6$) \\[2pt]
\hline
\rule{0pt}{3.8mm}
% N_He   ESMF     N5-ESMP2   N7-ESMP2   EOM-CCSD    CISD
   0   & 7.7286 &  8.4508  &  8.4353  &  8.1946  & 10.1593 \\ 
   1   & 7.7286 &  8.4508  &  8.4353  &  8.1946  & 10.5369 \\ 
   2   & 7.7286 &  8.4508  &  8.4353  &  8.1946  & 10.9118 \\ 
   3   & 7.7286 &  8.4508  &  8.4353  &  8.1946  & 11.2841 \\ 
   4   & 7.7286 &  8.4508  &  8.4353  &  8.1946  & 11.6537 \\ 
   5   & 7.7286 &  8.4508  &  8.4353  &  8.1946  & 12.0207 \\ 
   6   & 7.7286 &  8.4508  &  8.4353  &  8.1946  & 12.3852 \\ 
%%%%%%%%%%%%%%%%%%%%%%%%%%%%%%%%%%%%%%%%%%%%%%%%%%%%%%%%%%%%%%%%%%%%%%%%%%%
%%%%%%%%%%%%%%%%%%%%%%%%%% Data with more digits %%%%%%%%%%%%%%%%%%%%%%%%%%
%%%%%%%%%%%%%%%%%%%%%%%%%%%%%%%%%%%%%%%%%%%%%%%%%%%%%%%%%%%%%%%%%%%%%%%%%%%
%%
%%              ESMF       N7-ESMP2        CISD
%%        0    7.7286    ???????????    10.159298
%%        1    7.7286    ???????????    10.536938
%%        2    7.7286    ???????????    10.911840
%%        3    7.7286    ???????????    11.284060
%%        4    7.7286    ???????????    11.653661
%%        5    7.7286    ???????????    12.020690
%%        6    7.7286    ???????????    12.385205
%%
%%%%%%%%%%%%%%%%%%%%%%%%%%%%%%%%%%%%%%%%%%%%%%%%%%%%%%%%%%%%%%%%%%%%%%%%%%%
%%%%%%%%%%%%%%%%%%%%%%%%%%%%%%%%%%%%%%%%%%%%%%%%%%%%%%%%%%%%%%%%%%%%%%%%%%%
\hline\hline
\end{tabular}
\end{center}
\end{table}

\subsection{Size Intensive Excitation Energies}
\label{sec:size_intensivity}

Finally, although ESMP2 is rigorously size intensive
--- by which we mean that the excitation energy is unchanged
by adding a second, infinitely-far-away system that does not
participate in the excitation ---
it is worth testing that this property has been realized 
in our implementation.
To this end, we treated a water molecule with various numbers of
far-away helium atoms in a 6-31G basis.
We performed seven calculations, one with just the water molecule
and then six more, each with one additional He atom placed 10 \AA \ 
away from the water at the different points of an octahedron.
As seen in Table \ref{tab:size_test}, the ESMP2 prediction for
the excitation energy was unchanged by the addition of the
He atoms, both for the original N$^7$-scaling approach
and the N$^5$-scaling approach introduced here in which only
the dominant TOP is considered large.
While ESMP2's size intensivity is a formal advantage over CASPT2,
which is only approximately size consistent,
\cite{van1999}
CASPT2's size intensivity error turns out to be less than
$10^{-6}$ eV in this example.
In contrast, the excitation energy of configuration interaction
with singles and doubles (CISD),
which is not even approximately size consistent or intensive, changes
significantly
upon adding the He atoms, despite the fact that they have essentially
no interaction with the water molecule.
This alarming behavior is a reminder of why size-intensivity
is such a high priority in excited state methods, as artificial
energy shifts of the size displayed here by CISD could spoil 
predictions of solvation properties such as solvatochromic shifts.

%\begin{table}[t]
%\caption{Excitation energies (eV) for the TensorFlow and FastFock implementations. 
%         \label{tab:ee_compare}
%}
%\begin{tabular}{c c c}
%\hline\hline
%System & TensorFlow & Fast Fock \\
%\hline
%Cl$^-$-H$_2$O & 4.7195 & 4.7195 \\
%NH$_3$-F$_2$ & 4.5367 & 4.5367 \\
%\hline\hline
%\vspace{2mm}
%\end{tabular}
%\end{table}

\section{Conclusion}
\label{sec:conclusion}

We have shown that, by working in an orbital basis similar to that of
the natural transition orbitals and by making a small modification to the
zeroth order Hamiltonian, the cost scaling of the ESMP2 correction to
the ESMF energy can be lowered from the seventh to the fifth power
of the system size.
In particular, the scaling matches the $N_o^2N_v^3$
scaling of ground state MP2 theory, although the prefactor remains
significantly higher due to the off-diagonal nature of ESMP2's
zeroth order Hamiltonian, which necessitates an iterative solution
to the central linear equation.
Initial testing of this lower-scaling incarnation of ESMP2 theory shows that
its accuracy remains competitive with EOM-CCSD in many scenarios, but that
it may break down more rapidly when doubly excited character is present.
Given that this approach to ESMP2 gives it a lower cost-scaling than
EOM-CCSD, these findings strongly motivate more systematic and widespread
testing in future.
The potential for a low-scaling method that is robust in charge transfer
contexts is especially strong, as DFT still struggles in this area and
modeling these systems reliably often
requires the explicit inclusion of solvent species and can thus easily
entail hundreds of atoms.

Going forward, the immediate priority is to work towards a production-level
implementation of the most expensive terms within the theory.
Happily, our automatic code-generation and cost-analysis has revealed that
the number of terms with fifth order scaling is relatively small, and so
a hand-tuned implementation employing dense linear algebra should be
quite feasible.
Once the practical efficiency of the implementation is addressed, it will
be important to test the method in a significantly larger and more systematic
set of excitations in order to more firmly establish in which contexts
ESMP2 can be used as a lower-cost alternative to EOM-CCSD and in which
contexts it cannot.
Looking a bit farther ahead, it would be interesting to further exploit
locality.
The new approach here derives its scaling from the fact that molecular
excitations' spatial extents typically do not grow indefinitely with
system size, but it does not exploit localities of electron correlation
in the way many ground state methods now do.
Finally, the realization of an excited state analogue of MP2 theory
at the same cost scaling further motivates the study of applying a cluster
operator to the ESMF reference wave function, which would be an important
step towards the type of systematically improvable hierarchy of
correlation methods that Hartree Fock theory has long enjoyed.

%\begin{suppinfo} 

\begin{acknowledgments}
This work was supported by the National Science Foundation's
CAREER program under Award Number 1848012.
The Berkeley Research Computing Savio cluster performed the calculations.

J.A.R.S. acknowledges that this material is based upon work supported by
the National Science Foundation Graduate Research Fellowship Program under
Grant No. DGE 1752814.
Any opinions, findings, and conclusions or recommendations expressed in this
material are those of the author(s) and do not necessarily reflect the views
of the National Science Foundation.
\end{acknowledgments}

\bibliographystyle{achemso}
\bibliography{main} % Produces the bibliography via BibTeX.

%\appendix
%
%$\quad$
%
%Appendix goes here.

\end{document}